\documentclass[conference]{IEEEtran}

\usepackage{hyperref}
\usepackage{listings}
\usepackage{amsmath,amssymb,amsfonts}
\usepackage{algorithmic}
\usepackage{graphicx}
\usepackage{xcolor}
\usepackage{balance}

\hypersetup{
	colorlinks   = true, 
	urlcolor     = blue, 
	linkcolor    = blue, 
	citecolor   = red 
}

\lstdefinestyle{customc}{
	belowcaptionskip=1\baselineskip,
	breaklines=true,
	frame=L,
	xleftmargin=\parindent,
	language=C,
	showstringspaces=false,
	basicstyle=\footnotesize\ttfamily,
	keywordstyle=\bfseries\color{green!40!black},
	commentstyle=\itshape\color{purple!40!black},
	identifierstyle=\color{black},
	stringstyle=\color{orange},
}
\title{UEFI virtual machine firmware hardening through snapshots and attack surface reduction}
\providecommand{\keywords}[1]{\textbf{\textit{Index terms---}} #1}

\begin{document}
 \author{\IEEEauthorblockN{Mikhail Krichanov}
	\IEEEauthorblockA{ 
		\textit{Ivannikov Institute for System Programming}\\
		\textit{of the Russian Academy of Sciences}\\
		Moscow, Russia\\
		krichanov@ispras.ru}
	\and
	\IEEEauthorblockN{Vitaly Cheptsov}
	\IEEEauthorblockA{
		\textit{Ivannikov Institute for System Programming}\\
		\textit{of the Russian Academy of Sciences}\\
		Moscow, Russia\\
		cheptsov@ispras.ru}}
	
\maketitle
\thispagestyle{plain}
\pagestyle{plain}

\begin{abstract}
	
The Unified Extensible Firmware Interface (UEFI) is a standardised interface between the firmware and the operating system used in all x86-based platforms over the past ten years. A side effect of the transition from conventional BIOS implementations to more complex and flexible implementations based on the UEFI was that it became easier for the malware to target BIOS in a widespread fashion, as these BIOS implementations are based on a common specification. This paper introduces Amaranth project - a solution to some of the contemporary security issues related to UEFI firmware. In this work we focused our attention on virtual machines as it allowed us to simplify the development of secure UEFI firmware. Security hardening of our firmware is achieved through several techniques, the most important of which are an operating system integrity checking mechanism (through snapshots) and overall firmware size reduction.

\end{abstract}


\section{Introduction}

It has become common sense to model computer architecture as a multi-layer system. When describing a computing system on a high level, NIST defines four layers: application software, operating system, system firmware, and hardware \cite{NIST147, NIST193}. Here the combination of system firmware and hardware can comprise something that we call platform. The existence of a platform is meant to leverage the complexity of hardware initialisation tasks to the earlier boot stage in order to essentially simplify the operating system code.

The Unified Extensible Firmware Interface (UEFI) \cite{UEFI} is a standardised interface between the system firmware and the operating system used in all x86-based platforms over the past ten years. Migrating to the UEFI solved multiple operating system design issues, including the unification of the boot code, deprecation and removal of the undocumented vendor-specific extensions, interface simplification for early Input/Output (IO), boot path determination, hardware enumeration, and many more. When speaking about the UEFI, one generally considers three things:

\begin{itemize}
\item the UEFI firmware;
\item the UEFI specification;
\item the UEFI operating system.
\end{itemize}

The UEFI firmware, commonly named the UEFI BIOS (Basic Input/Output System) on x86 platforms for historic reasons, is the main component of system firmware. The UEFI operating system runs on top of the UEFI firmware and can usually be viewed as a tandem of a UEFI application, serving as an operating system bootloader, and a UEFI-compatible kernel, which is aware of the features brought by the UEFI firmware through the UEFI specification.

The malicious code running at the firmware level could be used to compromise any components that are loaded later in the boot process, including boot loader, hypervisor, and operating system. Compared to the legacy x86 firmware, which did not employ any built-in security features in the majority of the platforms, the UEFI firmware is supposed to build a chain of trust and ensure that the operating system it runs is not compromised. This change made the x86 system firmware an even more attractive target for the attacks, as it is now required to hijack the chain either on the operating system side or the firmware side. A side effect of the transition from conventional BIOS implementations to more complex and flexible implementations based on the UEFI was that it became easier for the malware to target BIOS in a widespread fashion, as these BIOS implementations are based on a common specification.

The industry community does pay attention to platform security issues, such as protecting the integrity of the firmware and mechanisms for updating it \cite{NIST147, NIST193}, but, as this paper shows, much still remains to be done in this field.
In section II we briefly discuss the latest industry advancements in defending the x86 platforms with the firmware. In section III we demonstrate how this knowledge applies to hardening the virtual machine firmware, and in section IV we provide the excerpts of our practical experience, gained while implementing the Amaranth project, designed to make our virtual machines more secure.

\section{State of the art}

Nowadays UEFI firmware from different vendors are extremely overcomplicated and tend to provide as many means of
interaction with the user as operating systems do. According to the recent studies \cite{DECAF}, optional code
(code, which does not participate in the boot process) can constitute up to 70\% (table \ref{table:Decaf}) of the total size of system firmware and can be safely removed without any impact on the functioning of the target applications.

\begin{table}[h!]
	\caption{Percentage of unnecessary code in modern firmware \cite{DECAF}.}
	\begin{center}
		\begin{tabular}{|l|c|c|c|} 
			\hline
			 & Original & Final & \\
			 \multicolumn{1}{|c|}{Motherboard} & firmware & firmware &  Reduction\\
			 & size (KiB) & size (KiB) & \\
			\hline
			SuperMicro A1SAi-2550F (V519) & 3013 & 903 & 70.91\% \\
			\hline
			Tyan 5533V101 & 4520 & 1916 & 39.82\% \\
			\hline
			HP DL380 Gen10 & 46102 & 27809 & 39.68\% \\
			\hline
			SuperMicro A1SAi-2550F (V827) & 3000 & 2108 & 29.76\% \\
			\hline
			SuperMicro A2SDi-12C-HLN4F & 3618 & 2680 & 25.91\% \\
			\hline
			SuperMicro A2SDi-H-TP4F & 3645 & 2766 & 24.12\% \\
			\hline
			SuperMicro X10SDV-8C-TLN4F & 4519 & 4209 & 6.87\% \\
			\hline
		\end{tabular}
	\label{table:Decaf}
	\end{center}
\end{table}

Optional firmware code is not only a waste of the SPI (Serial Peripheral Interface) flash memory,
it also has major security implications.
A recent study has shown that because of numerous additional modules in UEFI images, 
and large amount of code reuse between images, certain attacks can be easily and reliably automated \cite{Voodoo}.

The following are typical ways to infect UEFI firmware with a persistent rootkit or implant:

\begin{enumerate}
	\item {\bf Modifying an unsigned UEFI Option ROM}\\
    An Option ROM is PCI/PCIe expansion firmware (Peripheral Component Interconnect (Express) --
    I/O bus for connecting peripherals to the computer's motherboard) in x86 code located on a PCI-compatible device.
    An Option ROM is loaded, configured, and executed during the boot process. In 2012 a variety of techniques for infecting Apple laptops was introduced, including through Option ROMs \cite{Loukas}. At Black Hat 2015  an attack, named Thunderstrike, was demonstrated that infiltrated the Apple Ethernet adapter with modified firmware that loaded malicious code \cite{Thunder}.
    Specifically, Thunderstrike loaded the original Option ROM driver with additional code that was then executed because the firmware did not authenticate the Option ROM’s extension driver during the boot process.
    
	\item {\bf Adding/modifying a DXE driver}\\
    By modifying a legitimate DXE (Driver eXecution Environment) driver in the firmware, an attacker is able to introduce malicious code that will be executed in the preboot environment, at the DXE stage.
    One way to modify a DXE driver in the UEFI firmware image is to bypass the SPI flash protection bits by
    exploiting a privilege escalation vulnerability \cite{Cache,Rootkit}. 
    Elevated privileges allow the attacker to disable SPI flash protection by turning off the protection bits. 
    Another way is to exploit a vulnerability in the BIOS update process \cite{Chronomancy,SignedBIOS} 
    that allows an attacker to bypass update authentication and write malicious code to SPI flash memory.
    
	\item {\bf Adding a new bootloader (bootkit.efi)}\\ 
    An attacker can add another bootloader (or replace the old one) to the list of the available bootloaders by modifying the
    \texttt{BootOrder} or \texttt{Boot\#\#\#\#} EFI-variables, which determine the order of OS bootloaders \cite{Setup}.
\end{enumerate}

Summarizing the above, upon taking a random x86-based platform on the market, one will face a list of common security issues related to x86-firmware we outlined below.

\begin{enumerate}
	\item A bloated firmware codebase leads to an increase of the overall vulnerability surface and exploit availability.
	\item Unauthenticated BIOS updates lead to the installation of the UEFI implants, which compromise the entire system.
	\item Outdated BIOSes with known security issues \cite{LOJAX} also lead to an increase of availability of exploits.
	\item The UEFI bootloaders commonly fail to check all the files they use in the boot process, providing room to compromise the target system and bypassing the trust chains and security mechanisms such as UEFI Secure Boot \cite{Hole}.
	\item The UEFI Secure Boot is both extremely overcomplicated and lacking functionality. It misses certificate chains, which resulted in all Option ROM and third-party operating system vendors signing their code by the same certificate. At the same time the revocation only works on per-file or per-certificate basis, completely ignoring the sequential nature of vulnerability fixing, making rollback protection, generally solved by security counters, nearly impossible on x86. To add more, a convoluted signature format, requiring the file to be heavily parsed for the verification, contributes to the overall availability of implementation exploits.
	\item Misconfigured SPI flash protection mechanisms (registers BIOS\_CNTL, PRx \cite{JTAG}) allow an attacker to write malicious code
	to SPI flash memory and completely compromise an entire system.
	\item SMM: vulnerabilities in SMI handlers \cite{Sandman, Win8, CPU} lead to privilege escalation, i.e. execution of malicious code
	with SMM privileges.
	\item Malicious peripherals: unsigned Option ROMs and lack of separation of the degrees of trust for different devices, whether it's an embedded media or an external one, lead to DMA (Direct Memory Access) attacks \cite{CVE}.
	\item S3 resume boot-script \cite{Attack, Cr4sh} vulnerabilities are similar to those in SMI handlers as they lead to privilege escalation in PEI (Pre-Efi Initialisation) stage right after platform wake-up.
	\item Privileged coprocessors, such as Intel ME/AMT, are additional devices, which vulnerabilities allow to bypass any other security mechanisms of a system
	\cite{ME11, AMT}.
\end{enumerate}

Good firmware implement mitigations on both software and hardware levels to reduce the impact of the attacks belonging to each of these categories.

\section{Hardening firmware for virtual machines}

The development of a secure firmware is a vast problem. To narrow this task we decided to focus our attention on a simpler problem of 
creating a secure firmware for the virtual machines.
During the development of the firmware we had to keep in mind that the standard operating mode for the virtual machines is
rather atypical for physical machines owned by ordinary users. In such a mode the system does not sleep and rarely reboots.
When working in this mode, there is no need to emulate sleep states,
which can be disabled in the settings of the virtual machine. 
This measure allows us to reduce security threats coming from the S3 resume boot-script.

The absence of additional devices (such as Intel ME / AMT) for virtual machines or 
the ability to disable them in the settings (as in the case of Option ROM) also simplifies the task of developing a secure firmware and reduces the threat of DMA attacks.

In the settings of the virtual machine, one can disable the ability to write to SPI flash
memory and thereby get rid of the need to configure flash memory protection mechanisms (registers BIOS\_CNTL, PRx). Such configuration of a virtual machine substantially weakens the probability of implants in the UEFI firmware.

We could also turn off the SMM support in the virtual machine configurations and, consequently, reduce the security threats caused by the SMM, as we developed our own OS integrity checking mechanism, which we used instead of the UEFI Secure Boot. The UEFI Secure Boot is the only reason one would need to emulate the SMM in virtual machines and as we managed to dispose the UEFI Secure Boot, we could also leave out the SMM.

\begin{figure}[h]
	\centerline{\includegraphics[scale=0.6]{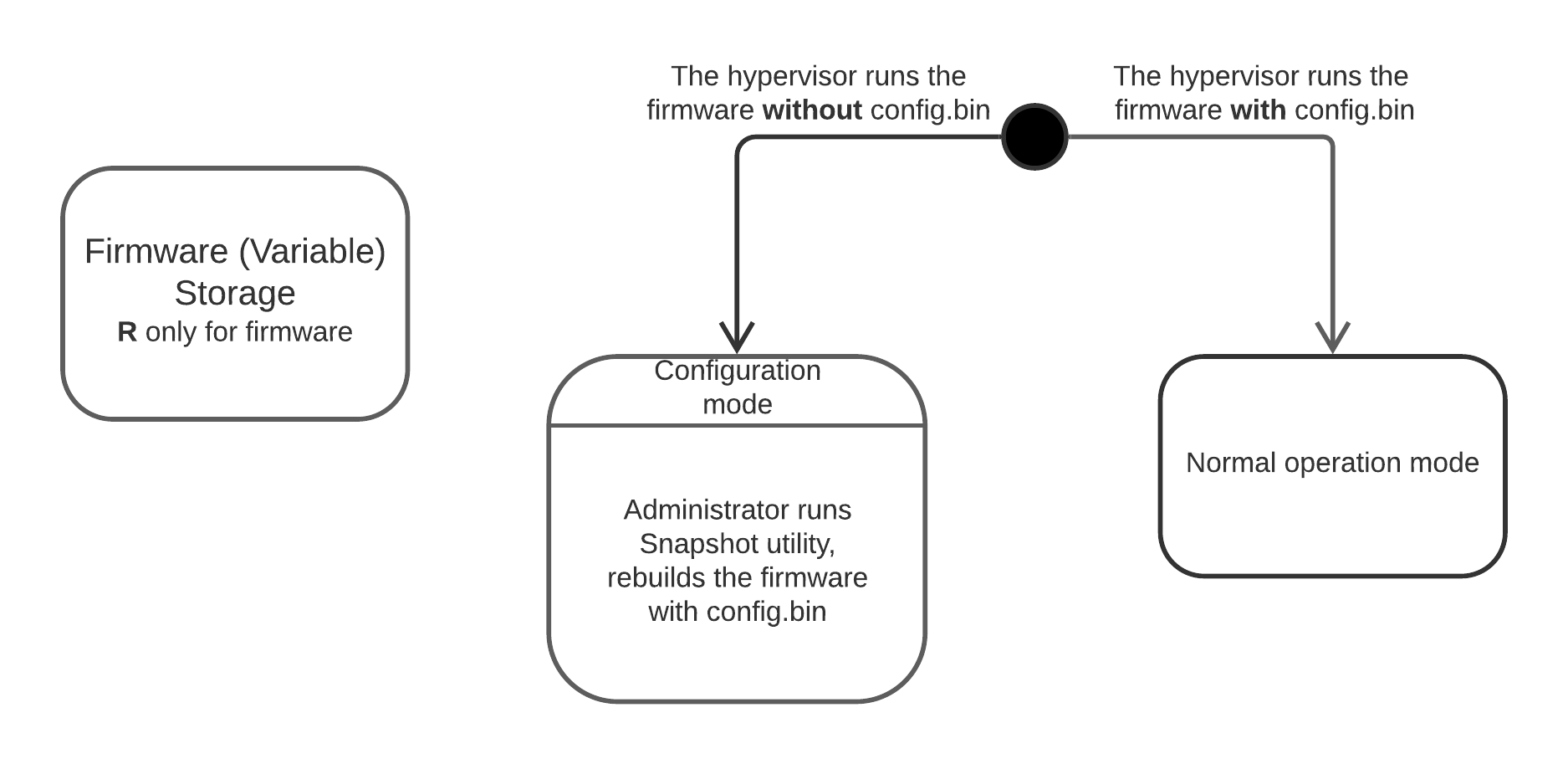}}
	\caption{Virtual machine flash storage structure corresponding to the firmware state machine.}
	\label{fig:Qemu}
\end{figure}

Furthermore, the firmware update responsibility is handed over to the hypervisor, running these machines. Taking into account all these simplifications, the resulting firmware state machine can look like one, depicted in Fig. \ref{fig:Qemu}.

We started the development of a secure firmware from the minimisation of the interface between the firmware and the user (both a person and an operating system). 
To do so, we took the open-source reference implementation of the UEFI firmware for virtual machines -- OVMF \cite{EDK2} and
removed several redundant features from its code such as a complex graphical user interface, network transfer protocol support, as well as various unsafe ways of finding and restoring the operating system (for example, booting from a USB flash drive). 
As a replacement for the UEFI Secure Boot, we developed a prototype of an operating system integrity checking mechanism, named Snapshot, which became the key feature of our {\bf Amaranth} project.

The Snapshot mechanism works similar to the Apple Secure Boot manifests, which represent a configuration file containing: 
checksums of the files of critical importance to the boot process of an operating system, the minimum version of the platform API 
(if the firmware is too old, boot is prohibited), a set of rules that impose restrictions on the contents of some directories.

While Apple's manifest is a file that is stored together with the bootloader, 
in case of Amaranth, {\bf configuration file} becomes part of the firmware. Unlike Apple,
we also cannot guarantee that the Windows or Linux bootloader will not try to use any unverified files, which means we have to sign more files (calculate their checksums) and check directories for the absence of files of certain types in them,
since loaders most often do not check anyhow the code they load. 
If the checksums do not match or directory contents are inappropriate, the system will not be allowed to boot. As long as the configuration file is managed by the host operating system and therefore comes from a trusted source, there is no need to implement any signature checking. However, if this mechanism is ported to real hardware, the situation will have to change.

This approach could have been made better if the operating systems cooperated with the process, for example, by using the standard UEFI mechanisms to access the file system. This way the requests could be intercepted and checked whether the file is present in the configuration and thus trusted. Unfortunately it is not the case for the official Windows bootloader, which uses its own NTFS-driver, 
forcing us to run additional checks in the firmware. While this is not the only possible solution as we could have used some other Windows loader (for example, Quibble \cite{Quibble}), 
which file system requests could be intercepted in a standard way, ensuring the security and compatibility of such a loader is non-trivial.

Apart from the need to maintain data integrity from the external code, implemented with the Snapshot mechanism, it should be understood that there is a need to protect a fair amount of code in the firmware itself, which comes from different parties. Obviously, all modern development tools for mission-critical software should always be used during the development of such a product. We performed formal verification of the PE loader with the AstraVer toolset \cite{PE}, fuzzed critical parsing code parts, such as firmware configuration handling code, and ran static analysis with Svace \cite{Svace}. To further reduce the risks of bugs residing in the codebase without our knowledge we created a hardened operating environment with the support for memory protection, stack canaries, stack overflow guards, and automatic variable initialisation.

\section{Snapshot mechanism architecture}

Before we describe the main security mechanism of Amaranth firmware in greater detail, it seems reasonable to explain the necessity of it in comparison with other integrity checking mechanisms. It must be noted that any attempt to pass the responsibility of checking the integrity of the booted OS to the OS itself breaks the security chain, as only the code we already trust can check the code we do not trust yet. For this reason UEFI Secure Boot was created. This technology is supposed to authenticate UEFI images, such as an OS loader or an option ROM and it is the only link in a chain of trust. Thus in case of UEFI Secure Boot the overall security of the system depends on the OS loader, and if it can not be trusted (section II, 4\textsuperscript{th} security issue), one will have to sign all the system critical files and store them in SPI flash memory along with the firmware, using mechanisms provided by UEFI Secure Boot. Not taking into the account other design decisions (section II, 5\textsuperscript{th} security issue) UEFI Secure Boot was designed to authenticate several bootloaders and option ROMs, not a huge amount of system files, what resulted in overhead of 24 bytes per entry ({\it WIN\_CERTIFICATE}), where certificate type and hash algorithm must be specified.
These certificates are stored as authenticated UEFI variables in the SPI flash memory, thus the correctness of UEFI Secure Boot depends on the SPI flash memory protection mechanisms (section II, 6\textsuperscript{th} security issue) and on the signature databases updating procedures. Section 32.5.1 of UEFI specification \cite{UEFI} states that ``signature databases may be updated by the 
firmware, by a pre-OS application or by an OS application or driver''. But leaving signature databases updates to guest OS application or driver can hardly be regarded as a wise decision (section II, 2\textsuperscript{nd} security issue). Nevertheless even if these databases are updated by firmware, it turns out that firmware from  different vendors broadly reuse code from open source reference UEFI implementation \cite{DECAF} and that the implementation (OVMF \cite{EDK2}) uses SMM in UEFI variable driver stack (section II, 7\textsuperscript{th} security issue). Taking into account all the aforementioned considerations we decided to develop our own integrity checking mechanism (Snapshot).

Checking disk partitions and specific directories within them for the absence (presence) of files of a certain type is implemented in Amaranth with the help of a set of entries stored in a configuration file. 
Each entry includes a full path to the base directory, a set of regular expressions describing the contents of the base directory,
and a flag indicating whether the set of files described by the regular expressions is valid or not.

A small utility named {\bf Snapshot} was written to create the firmware configuration files. This program takes text files for input, with each file containing a set of full paths to critical files within a single disk partition or a set of rules describing the contents of required directories within the same partition. Upon obtaining all the necessary information, the utility generates a binary file, which the user can then place into the firmware and secure the operating system state at boot time. To create a configuration file, the Snapshot utility needs multiple sets of 4 arguments, each one describing a single disk partition. These arguments are:
\begin{enumerate}
	\item Path to a file (partN\_files.txt) containing a set of lines with full file paths, which hashes are to be computed.
	\item Partition N type Globally Unique Identifier (PartitionTypeGUID).
	\item Partition N Globally Unique Identifier (UniquePartitionGUID).
	\item Path to a file (partN\_rules.txt) containing a set of rules describing the partition N contents. An example of such a description is shown in Listing \ref{lst:rules}.
\end{enumerate}

\begin{lstlisting}[caption={An example of partN\_rules.txt file.},label={lst:rules}]
#WR
C:\Program Files\Common Files\System\msadc
*.dll
*.inc
*.reg
ru-RU\*.dll.mui
#BN
C:\Windows\Boot\PCAT
DtcInstall.log
\end{lstlisting}

This file describes the contents of the partition by listing valid or invalid file names relatively to the base directory. The full path to the base directory is specified in the line following the line of flags. A line of flags begins with the special character \#,
followed by a sequence of letters describing the contents of the base directory. There are 4 allowed flags:
\begin{itemize}
	\item W -- indicates a list of valid files (whitelist).
	\item B -- indicates a list of invalid files (blacklist).
	\item R -- the lines following the line of the base directory up to the next line of flags or the end of the file are regular expressions (written relative to the base directory).
	\item N -- the lines following the line of the base directory up to the next line of flags or the end of the file are ordinary file names (written relative to the base directory).
\end{itemize}

The contents of each base directory can be described only once. The regular expression syntax is extremely simple and contains only 2 control sequences:
\begin{itemize}
	\item ? -- denotes one arbitrary but valid character;
	\item $\star$ -- denotes an arbitrary number of such characters.
\end{itemize} 
Valid characters are those that can be used in file names.

When the platform starts, the compliance of the OS state with the configuration file, 
the structure of which is shown in Listing \ref{lst:struct}, is checked by a special library firmware module, 
which has been designed taking into account critical security requirements.
Despite the importance of these requirements, they are often neglected \cite{PE}.
When applied to the configuration file parser, the security requirements are quite simple:
\begin{enumerate}
	\item When one passes any structure to a function, one should always pass not only a pointer to this structure, but also its size. 
	This allows one to make elementary checks inside the function, that any pointers (i.e. offsets relative to the beginning of the
	binary file) inside this structure do not point outside the memory allocated for this structure.
	\item All address arithmetic operations must be performed taking into account the possibility of overflow. 
	If such a situation occurs, an error must be reported and the system must not be allowed to boot.
	\item Upon completion of its execution, each function must return the status of the operation it performed,
	so that no error could be left unnoticed.
	\item Files being checked for validity should be read into the buffer only once, 
	and all the checks should be performed over this buffer
	only, in order to eliminate the possibility of TOCTOU attacks (Time-of-check/Time-of-use).
\end{enumerate}

\begin{lstlisting}[caption={Configuration file structure},label={lst:struct}]
/**
Storage magic.
**/
#define SS_STORAGE_MAGIC SIGN_32 ('S', 'S', 'O', 'H')

/**
Storage version.
**/
#define SS_STORAGE_VERSION 0x10010000U

/**
No boot partition index.
**/
#define SS_BOOT_PARTITION_UNUSED MAX_UINT32

/**
File entry describing one file and its hash.

All the files are alphabetically-sorted,
thus binary searching can be used to access files.

Offset must point within the binary blob.
**/
typedef struct {
UINT32            Offset;
UINT8             Hash[SHA384_DIGEST_SIZE];
} SS_STORAGE_FILE;

/**
Rules described in ACL contain a whitelist
when the bit flag is set and blacklist 
when the bit flag is not set.
**/
#define ACL_FLAG_WHITELIST  BIT0

/**
Rules described in ACL contain regular expressions
instead of raw strings. Currently supported format:
- ? stands for any ASCII character.
- * stands for any amount of ASCII characters.
**/
#define ACL_FLAG_REGEX      BIT1

/**
Access control rule is a rule describing 
restrictions imposed on a directory 
referred by the rule. The rule can be viewed
as a way to whitelist or blacklist directory 
contents without explicitly checking file contents.

This can be used to protect directories from 
having new malicious files added in blacklist
or whitelist modes:
- Blacklist mode forbids the files 
  that can be matched by any rule
- Whitelist mode restricts the files to only those
  that can be matched by any rule.
**/
typedef struct {
UINT32    Flags;
UINT32    DirectoryOffset;
UINT32    NumberOfRules;
UINT32    Rules[];
} SS_STORAGE_ACL;

/**
Storage partition describing one storage and
its files.
**/
typedef struct {
EFI_GUID          PartitionTypeGUID;
EFI_GUID          UniquePartitionGUID;
UINT32            NumberOfAclRules;
UINT32            AclRuleOffset;
UINT32            NumberOfFiles;
SS_STORAGE_FILE   Files[];
} SS_STORAGE_PARTITION;

/**
Storage set describing all the partitions
in the configuration.
**/
typedef struct {
UINT32             Magic;
UINT32             Version;
UINT32             BootPartitionIndex;
UINT32             BooterFileOffset;
UINT32             NumberOfPartitions;
UINT32             Partitions[];
} SS_STORAGE_SET;
\end{lstlisting}

The configured firmware runs as follows: after the successful completion of the SEC, PEI and DXE stages, 
the firmware proceeds to the execution of the stage called Boot Device Selection (BDS),
at the end of which the {\it BdsSnapshotDeviceSelect()} function is called, which passes the control to the configuration file parser.
First of all, the parser checks, that the given file is indeed a configuration file by comparing the values of the {\it Magic} and 
{\it Version} fields of the {\it SS\_STORAGE\_SET} structure with the expected values. 
Further, the parser makes sure that the index of the disk partition in which the OS loader is stored 
({\it BootPartitionIndex}) does not exceed the total number of partitions ({\it NumberOfPartitions}). 

On the next step the parser checks each disk partition. Information about partitions is stored in the configuration file as an array of
{\it SS\_STORAGE\_PARTITION} structures. At first the parser checks, that different partitions on the disk do not have the same non-zero
identifiers ({\it UniquePartitionGUID}). Then it compares the contents of each partition with specifications in the configuration file,
namely: the parser makes sure, that the partition type matches the expected one ({\it PartitionTypeGUID});
after that the parser compares checksums of each of {\it NumberOfFiles} files on the disk with those stored in configuration file 
as an array of {\it SS\_STORAGE\_FILE} structures, each containing 2 fields: 
\begin{itemize}
	\item offset from the beginning of the configuration file, 
    which points at a string containing a full file name (+ end-of-line character);
    \item a checksum of the file ({\it Hash[SHA384\_DIGEST\_SIZE]}).
\end{itemize}

On the third step, the entire file system of the disk partition is checked for compliance with each of {\it NumberOfAclRules} rules,
which impose restrictions on the types of files located in certain directories, and which are stored in the configuration file as an
array of {\it SS\_STORAGE\_ACL} structures. Each of these structures contains a set of fields: 
\begin{itemize}
	\item an offset from the beginning of the configuration file ({\it DirectoryOffset}), which points at a line with a full name of the
	directory (+ end-of-line character), on the contents of which restrictions are imposed; 
	\item an array of {\it NumberOfRules} lines, that are stored in the configuration file at the offsets specified in {\it Rules[~]};
	\item another field ({\it Flags}) contains a set of flags, which allows to interpret this array of strings.
\end{itemize}
Lines in the {\it Rules[~]} array can contain either relative file names or regular expressions. Both file names and regular expressions
are written relative to the base directory ({\it DirectoryOffset}). What exactly is stored in the {\it Rules[~]} lines is determined by
the {\it ACL\_FLAG\_REGEX} flag. The {\it ACL\_FLAG\_WHITELIST} flag allows the parser to determine,
whether filenames matching the {\it Rules[~]} are valid for the base directory or, on the opposite, not allowed.

If all the checks complete successfully, the parser passes the control to the bootloader, which location is determined by {\it BootPartitionIndex} and a full file name ({\it BooterFileOffset}) within this partition. 
If any discrepancy is found between the contents of the disk and the configuration file, 
an error is reported and the system is not allowed to start.

As it was mentioned earlier, the configuration file contains checksums of files, that are important to the operating system boot process. 
As there can be quite a lot of such files, it is reasonable to speed up the process of checksum calculations in the firmware.
For this purpose, the SHA384 algorithm was implemented in the firmware, which uses the AVX (Advanced Vector Extensions) command extension
for x86 platforms to accelerate calculations at the hardware level \cite{SHA512,SHA384}. 
Acceleration is achieved as follows:
\begin{itemize}
	\item {\it Xmm} registers are used, with which one can perform operations on two 8-byte words at a time.
	\item During checksum calculations cyclic permutations of 8-byte words of the intermediate result are done.
	In order not to copy words from one register to another, {\it virtual registers} are used instead
	(registers' aliases are provided to the assembler instructions), what allows us to simply rename them and spare some instructions.
	\item Since the basic calculations of the SHA384 algorithm can be split into two subtasks, one of which is performed
	exclusively on {\it Xmm} registers, and the other -- exclusively on general-purpose registers; to speed up the algorithm as a whole, 
	{\it Xmm} instructions can be interleaved with general-purpose instructions in a ratio of 1 to 2 and thus, one can
	take advantage of the hardware parallelism of execution of vector instructions and general-purpose instructions inside Intel
	processors (because on average it takes 2 clock cycles to execute a single vector instruction). 
\end{itemize}

\begin{figure}[h]
	\centerline{\includegraphics[scale=0.6]{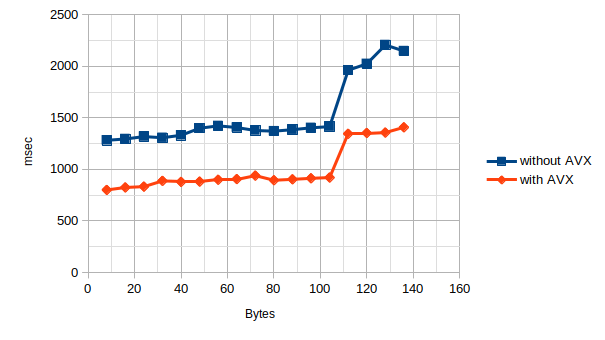}}
	\caption{Dependence of the checksum calculation time on the size of the message being hashed.}
	\label{fig:Avx}
\end{figure}

As can be seen in Fig. \ref{fig:Avx}, the use of aforementioned ideas allows to reduce the time of checksum calculation by 35\%. 
The measurements were made on one Intel (R) Core (TM) i5-3337U 1.80GHz processor (1 core), 
QEMU emulated the Standard PC (Q35 + ICH9, 2009) chipset with 2GiB RAM. The graph shows the time in milliseconds taken to compute one
million checksums for messages, lengths of which are given in bytes. 
The time increase at 112 bytes is due to the fact that the checksum is calculated for a padded message. 
On one of its steps the padding algorithm adds $k$ zeros to the original message. 
$k$ is determined as the smallest non-negative solution of the equation $l + 1 + k \equiv 896 \mod 1024$, where $l$ is the length of the original message in bits. Thus, if $l = 112 \times 8 = 896$, then $k = 1023$, which doubles the length of the message being hashed compared to the case when $l = 104 \times 8 = 832$ and $k = 63$.

In order to increase robustness of the configuration files firmware parser, fuzzing was conducted with the help of libFuzzer \cite{Fuzzer} library. 
The main difficulty in carrying out such testing was the preparation of a UEFI-compatible environment in the user space 
of the operating system, i.e. we needed functions, that would emulate the behavior of the platform on which the firmware was running.
So as to emulate the UEFI environment, a small utility from the OpenCore \cite{OpenCore} project was used, 
which simplified the preparations to fuzzing and reduced them to writing code, repeating the logics of the configuration file parser.

\section{Conclusion}

This paper introduced {\bf Amaranth}, a project which aims to develop a secure UEFI firmware for virtual machines.
The implemented solution for virtual machines running on the QEMU emulator includes:
\begin{itemize}
	\item a minimised interface between the firmware and the user (both a person and an operating system);
	\item a prototype of an operating system integrity checking mechanism implemented as a firmware library module based on the SHA384 hashing algorithm optimised with the AVX (Advanced Vector Extensions) command extensions for x86 platforms to accelerate calculations on the hardware level;
	\item a hardened operating environment that supports some of the latest security practices including memory protection, stack overflow guards, stack canaries, and automatic variable initialisation;
	\item the use of modern development tools for mission-critical software such as formal verification (PE loader \cite{PE}), 
	fuzzing (parser of firmware configuration files), and static analysis (Svace \cite{Svace});
	\item the support for booting both Windows (from 7 to 10) and Linux operating systems, with the ability to bypass the GRUB boot loader for the latter, on x86 and x86\_64;
	\item utility code to create firmware configuration files on all major desktop operating systems.
\end{itemize}

The proposed set of tools of the Amaranth project allows one to substantially reduce the following threats:
\begin{itemize}
	\item UEFI firmware implants, which may appear after software reboots of the OS, bypassing the hypervisor.
	\item OS bootloader modifications, as well as the alternation of files it depends on in the boot process (in the same reboot scenario). 
	\item Privilege escalation caused by the exploitation of the vulnerabilities in firmware runtime services (e.g. over SMM or through non-volatile variable access).
\end{itemize}

Designing all the security measures as a part of the UEFI firmware makes the approach highly portable. For this reason we plan to continue to develop Amaranth with the existing and new security measures for UEFI-compatible workstations and embedded devices as well.

\section*{Acknowledgements}

We would like to thank our colleagues in the ISP RAS Linux Verification Center, Alexey Khoroshilov and Evgeniy Baskov in particular, for their aid on the operating system side. We would also like to thank Marvin H{\"a}user from from University of Kaiserslautern for his incredible contributions into the EDK~II and our firmware.

\IEEEtriggeratref{9}

\end{document}